# Quantum oscillations and Fermi surface in an underdoped high temperature superconductor

Nicolas Doiron-Leyraud [1], Cyril Proust [2], David LeBoeuf [1], Julien Levallois [2], Jean-Baptiste Bonnemaison [1], Ruixing Liang [3,4], D.A. Bonn [3,4], W.N. Hardy [3,4] & Louis Taillefer [1,4]

*1 Département de physique & RQMP, Université de Sherbrooke, Sherbrooke, Canada*

*2 Laboratoire National des Champs Magnétiques Pulsés, UMR CNRS-UPS-INSA 5147, Toulouse, France*

*3 Department of Physics & Astronomy, University of British Columbia, Vancouver, Canada*

*4 Canadian Institute for Advanced Research*

**Despite twenty years of research, the phase diagram of high-temperature superconductors remains enigmatic[1],[2]. A central question is: what makes the physical properties of these copper oxides so dramatically different on opposite sides of the superconducting region?  In the overdoped regime, the material behaves as a reasonably conventional metal, with a large Fermi surface[3],[4].  The underdoped regime, however, is highly anomalous and appears to have no coherent Fermi surface, but only disconnected "Fermi arcs"[5],[6].  The fundamental question, then, is whether underdoped cuprates have a Fermi surface, and if so, whether it is topologically different from that seen in the overdoped regime.  Here we report the observation of quantum oscillations in the electrical resistance of the oxygen-ordered cuprate YBa$_2$Cu$_3$O$_{6.5}$, establishing the existence of a well-defined Fermi surface in the ground state of underdoped cuprates, once superconductivity is suppressed by a magnetic field.  The low oscillation frequency reveals a Fermi**

**surface made of small pockets, in contrast to the large cylinder characteristic of the overdoped regime. Two possible scenarios are discussed: either a small pocket is part of the band structure specific to YBa$_2$Cu$_3$O$_{6.5}$ or small pockets arise from a topological change at a critical point in the phase diagram. Our understanding of high-temperature superconductors will depend critically on which of these two scenarios is proven correct.**

The electrical resistance of two samples of ortho-II ordered YBa$_2$Cu$_3$O$_{6.5}$ was measured in a magnetic field up to 62 T applied normal to the CuO$_2$ planes ($B \parallel c$). (Sample characteristics and details of the measurements are given in the Methods section below.) With a $T_c$ of 57.5 K, these samples have a hole doping per planar copper atom of $p = 0.10$, *i.e.*, they are well into the underdoped region of the phase diagram (see Fig. 1a). ARPES data for underdoped Na$_{2-x}$Ca$_x$Cu$_2$O$_2$Cl$_2$ (Na-CCOC) at precisely the same doping (reproduced in Fig. 1b from ref. 6) shows most of the spectral intensity to be concentrated in a small region near the nodal position ($\pi/2$, $\pi/2$), suggesting a Fermi surface broken up into disconnected arcs, while ARPES studies on overdoped Tl$_2$Ba$_2$CuO$_{6+\delta}$ at $p = 0.25$ reveal a large, continuous cylinder (reproduced in Fig. 1c from ref. 4).

The Hall resistance $R_{xy}$ as a function of magnetic field is displayed in Fig. 2 for sample A, and in Fig. S1 for sample B, where oscillations are clearly seen above the resistive superconducting transition. Note that a vortex liquid phase is believed to extend well above the irreversibility field, beyond our highest field of 62 T, which may explain why $R_{xy}$ is negative at these low temperatures, as opposed to positive at temperatures above $T_c$. Nevertheless, quantum oscillations are known to exhibit the very same diagnostic characteristics of frequency in the vortex state as in the field-induced normal state above $H_{c2}(0)$ (*e.g.* ref. 7). They are caused by the passage of quantized Landau levels across the Fermi level as the applied magnetic field is varied, and as such they are considered the most robust and direct signature of a coherent Fermi surface (FS). The inset of Fig. 2 shows the 2-K isotherm and a smooth background curve. We extract the oscillatory component, plotted in Fig. 3a as a function of inverse field, by subtracting the monotonic background (shown for all temperatures in Fig. S2). This shows that the oscillations are periodic in $1/B$, as expected of oscillations that arise from Landau quantization. A Fourier transform yields the power spectrum, displayed in Fig. 3b, which consists of a single frequency, $F = 530 \pm 10$ T. In Fig. 3c, we plot the amplitude of the oscillations as a function of temperature, from which we deduce a carrier mass $m^* = 1.9 \pm 0.1$ $m_0$, where $m_0$ is the bare electron mass. Within error bars, both $F$ and $m^*$ are the same in sample B, for which $J \parallel b$ (see Fig. S1). Oscillations of the same frequency are also observed in $R_{xx}$ (in both samples), albeit with a smaller amplitude. We note that while at 7.5 K the oscillations are still perceptible, they are absent at 11 K, as expected from thermally damped quantum oscillations (see Fig. S5).

While quantum oscillations in $YBa_2Cu_3O_{6+y}$ (YBCO) have been the subject of a number of earlier studies[8,9,10], the data reported so far do not exhibit clear oscillations as a function of $1/B$ and, as such, have not been accepted as convincing evidence for a Fermi surface[11]. Furthermore, we note that all previous work was done on oriented powder samples as opposed to the high-quality single crystals used in the present study.

Quantum oscillations are a direct measure of the FS area via the Onsager relation: $F = ( \Phi_0 / 2\pi^2 ) A_k$, where $\Phi_0 = 2.07 \times 10^{-15}$ T $m^2$ is the flux quantum, and $A_k$ is the cross-sectional area of the FS normal to the applied field. A frequency of 530 T implies a Fermi surface pocket that encloses a $k$-space area (in the $ab$ plane) of $A_k = 5.1$ $nm^{-2}$, i.e. 1.9 % of the Brillouin zone (of area $4\pi^2 / ab$). This is only 3 % of the area of the FS cylinder measured in Tl-2201 (see Fig. 1c), whose radius is $k_F \approx 7$ $nm^{-1}$. In the remainder, we examine two scenarios to explain the dramatic difference between the small FS revealed by the low frequency of quantum oscillations reported here for YBa$_2$Cu$_3$O$_{6.5}$ and the large cylindrical surface observed in overdoped Tl$_2$Ba$_2$CuO$_{6+\delta}$. The first scenario assumes that the particular band structure of YBa$_2$Cu$_3$O$_{6.5}$ is different and supports a small FS sheet. In the second, the electronic structure of overdoped cuprates undergoes a transformation as the doping $p$ is reduced below a value $p_c$ associated with a critical point.

Band structure calculations for stoichiometric YBa$_2$Cu$_3$O$_{6+y}$ ($y = 1.0$), which is slightly overdoped (with $p = 0.2$), show a FS consisting of four sheets[12,13], as reproduced in Fig. 4a: two large cylinders derived from the CuO$_2$ bi-layer, one open surface coming from the CuO chains, and a small cylinder associated with both chain and plane states. The latter sheet, for example, could account for the low frequency reported here. ARPES studies on YBCO near optimal doping[14,15] appear to be in broad agreement with this electronic structure. However, recent band structure calculations[16] performed specifically for YBa$_2$Cu$_3$O$_{6.5}$, which take into account the unit cell doubling caused by the ortho-II order, give a FS where the small cylinder is absent, as shown in Fig. 4b. This leaves no obvious candidate FS sheet for the small orbit reported here.

The fact that the same oscillations are observed for currents along $a$ and $b$ suggests that they are not associated with open orbits in the chain-derived FS sheet. In YBCO, the CuO chains along the $b$-axis are an additional channel of conduction, responsible for an anisotropy in the zero-field resistivity $\rho(T)$ of the normal state (above $T_c$). In Fig. S3, we plot the anisotropy ratio $\rho_a / \rho_b$ as a function of temperature, and the chain resistivity vs $T^2$, obtained from samples A and B. Although the anisotropy is about 2.2 at 300 K, it eventually disappears as $T \rightarrow T_c$, with $\rho_a / \rho_b \approx 1.0$ for $T < 80$ K, due to the localization of carriers in the quasi-1D chains. We thus conclude that chains make a negligible contribution to electrical transport at the low temperatures where quantum oscillations are observed. The oscillations must therefore come from the electronic structure of the CuO$_2$ planes.

The apparent lack of agreement between our experimental data and the calculated band structure of ortho-II $YBa_2Cu_3O_{6.5}$ calls for a careful re-examination of the band structure of YBCO near $p = 0.1$, both with and without ortho-II order. Let us then consider the possibility that electron correlations and/or the onset of an ordered phase with broken symmetry bring about a radical change in the FS from that predicted by LDA calculations.

Much like in chromium where a spin-density-wave (SDW) order causes the FS to reconstruct, a transformation in the electronic structure of YBCO could take place as a function of doping, resulting in small FS pockets. Such a scenario was invoked to explain the electron-doped cuprates[17]. Comparison of $A_k$ in YBCO with ARPES measurements[6] on Na-CCOC at the same doping ($p = 0.1$) suggests that the small FS pockets observed here are centered at nodal positions, that is to say at the crossing of the original large FS cylinder with the diagonal from (0, 0) to ($\pi$, $\pi$), *i.e.,* near ($\pi/2$, $\pi/2$). In Fig. 1b, we show how an elliptical FS pocket with an area of 5 nm$^{-2}$ located at this position would indeed include most of the ARPES spectral weight. Extending this to the four nodal positions could be a solution to the intriguing problem of the "Fermi arcs", revealing them to be elongated Fermi pockets. Let us examine some implications of this four-pocket scenario.

Assuming that the FS is strictly 2D, *i.e.,* neglecting any *c*-axis dispersion, then the carrier density *n* per plane per area is given by Luttinger's theorem as $n = 2 A_k / (2\pi)^2 = F / \Phi_0$ for each pocket. For four pockets, this gives $n = 1.02$ nm$^{-2} = 0.152 \pm 0.003$ carriers per planar Cu atom ($ab = 0.38227 \times 0.38872$ nm$^2$ is the area per Cu atom). Thus we find that $n / p = 1.5$. Taken literally, this would imply that the addition of one hole doped into the $CuO_2$ plane does not result in identically one extra charge carrier or, in other words, the Luttinger sum rule that relates the number of mobile carriers to the Fermi surface volume (or area in 2D) appears to be violated. This intriguing possibility depends on a number of assumptions that will require careful examination. First, we assume that the Onsager relation applies in cuprates exactly as it does in normal metals, *i.e.,* it is the frequency of oscillations in inverse applied magnetic field, rather than some other internal field, which is related to the Fermi surface area via $F = (\Phi_0 / 2\pi^2) A_k$. Secondly, we ignore the possibility that electron FS pockets may also exist, although it is not clear why they were not seen here. Finally, we neglect any dispersion along the *c*-axis. However, in order to recover the Luttinger sum rule solely on this ground, a rather strong dispersion is necessary, something not predicted by band calculations nor evidenced by an additional quantum oscillation frequency.

For comparison, the carrier density in overdoped Tl-2201 at $p = 0.25$ is $n = 1.26 \pm 0.04$ hole per Cu atom, as obtained by Hall effect[18], AMRO[3], and ARPES[4]. In Fig. 1c, this is directly given by the measured FS area. This is precisely the value expected in the large-FS scenario of band theory, where $n = 1 + p$. Extending this down to $p = 0.1$ would predict that $n$ should be 1.1 in $YBa_2Cu_3O_{6.5}$. The fact that we get $n = 0.15$ when we assume that the FS consists of no other sheet than that manifest in the single frequency observed in our experiment means one of two things: either it is incorrect to extend the $n = 1 + p$ relation down into the underdoped regime, or large parts of the FS of YBCO were not seen in our experiment (possibly because of a much larger $m^*$).

In other words, in the four-nodal pocket scenario a transition between a large-FS metal (in the overdoped regime) and a small-FS metal (in the underdoped regime) occurs at a critical doping somewhere between $p = 0.1$ and $p = 0.25$. The associated drop in carrier density $n$ by one order of magnitude would be expected to show up in most physical properties, for example the superfluid density $n_S$ and the normal-state electronic specific heat coefficient $\gamma_N$. Assuming the dynamical mass $m_d^* \approx 2 m_0$, close to the thermal mass $m^*$ deduced from quantum oscillations, we get $n_S = m_d^* c^2 / (4 \pi e^2 \lambda^2) \approx 0.12$ carriers per planar Cu atom, from the experimentally measured in-plane penetration depth of ortho-II ordered YBCO ($\lambda_a$ = 200 ± 20 nm; ref. 19). This low value of $n_S$ is much closer to the carrier density obtained in the four-nodal-pocket scenario ($n$ = 0.15) than in the large-FS scenario ($n$ = 1+ $p$ = 1.1). Note, however, that a number of factors need to be considered carefully to make such a quantitative comparison compelling, including the real dynamical mass and the possibility of strong phase fluctuations. The loss in density of states that would follow a FS transformation into small pockets could also explain the large and sudden drop in the specific heat coefficient $\gamma_N$ that has been observed in various cuprates as the doping is reduced below $p \approx 0.19$ (ref. 20).

A number of theories predict a Fermi surface made of four small pockets at nodal positions in the underdoped regime, going over to a large FS when $p$ exceeds a critical value $p_c$. Some of these are analogous to the usual SDW scenario[17] in the sense that they invoke the onset of an ordered phase with broken symmetry below $p_c$ (refs. 21,22,23), while others do not require any broken symmetry[24,25,26].

In summary, our unambiguous observation of quantum oscillations in underdoped YBCO proves the existence of a Fermi surface. The small size of the FS pocket associated with the low oscillation frequency suggests two very different scenarios for the non-superconducting ground state of underdoped cuprates. The first is a multi-band scenario in which the ground state is described by the LDA band structure. In the second scenario, the pseudogap phase that lies to the left of $T^*$ in the phase diagram is a highly correlated electronic fluid with a Fermi surface made of small pockets at nodal positions, separated from the Fermi liquid of the overdoped regime by a critical point near optimal doping. Our understanding of the underlying mechanisms that govern the behaviour of electrons in high-temperature superconductors will depend in a fundamental way on which of these two scenarios is proven correct.

**Methods**

**Samples**. The samples used are fully detwinned crystals of $YBa_2Cu_3O_{6+y}$ grown in non-reactive $BaZrO_3$ crucibles from high-purity starting materials (see ref. 27). The oxygen content was set at $y = 0.51$ and the dopant oxygen atoms were made to order into an ortho-II superstructure of alternating full and empty $CuO_y$ chains, yielding a superconducting transition temperature $T_c = 57.5$ K. The samples are uncut, unpolished thin platelets, whose transport properties are measured via gold evaporated contacts (contact resistance $< 1 \Omega$), in a six-contact geometry. Sample A (current along $a$-axis) and sample B (current along $b$-axis) have dimensions $40 \times 520 \times 720$ $\mu m^3$ (thickness $\times$ length $\times$ width) and $65 \times 810 \times 1030$ $\mu m^3$, respectively.

**Estimates of hole doping**. The hole doping $p$ in YBCO is determined from a relationship between $T_c$ and the $c$-axis lattice constant (see ref. 28). For our samples, the measured $T_c = 57.5$ K implies $p = 0.099$ and the measured $c = 1.17441 \pm 0.00005$ nm gives $p = 0.098 \pm 0.001$ (ref. 28).

**Resistance measurements**. Longitudinal ($R_{xx}$) and transverse ($R_{xy}$) resistances are obtained from the voltage difference measured diagonally on either side of the sample width, for a field parallel (up) and anti-parallel (down) to the $c$-axis: $R_{xx} \equiv (V_{up} + V_{down})$ / $2I_x$ and $R_{xy} \equiv (V_{up} - V_{down})$ / $2I_x$. A current excitation of 5 mA at 40 kHz was used. The voltage (and a reference signal) was digitized using a high-speed digitizer and post-analyzed to perform the phase comparison. The measurements were performed at the LNCMP in Toulouse, in a pulsed resistive magnet up to 62 T (ref. 29). Data for the rise (26 ms) and fall (110 ms) of the field pulse were in perfect agreement, thus excluding any heating due to eddy current.

**Supplementary Information** is linked to the online version of the paper at www.nature.com/nature.

**Acknowledgements** We thank R.T. Brisson, G.G. Lonzarich, G.L.J.A. Rikken and A.-M.S. Tremblay for inspiring discussions, and M. Nardone and A. Audouard for their help with the experiment and analysis. We acknowledge support from the Canadian Institute for Advanced Research and the LNCMP, and funding from NSERC, FQRNT, and a Canada Research Chair. Part of this work was supported by the French ANR IceNET and EuroMagNET.

**Author Contributions** N.D.-L. and C.P. contributed equally to this work.

**Author Information** Reprints and permissions information is available at www.nature.com/reprints. The authors declare no competing financial interests. Correspondence and requests for materials should be addressed to C.P. (proust@lncmp.org) or L.T. (louis.taillefer@physique.usherbrooke.ca).

**Figure 1 | Phase diagram of high-temperature superconductors.**

**a**, Schematic doping dependence of the antiferromagnetic ($T_N$) and superconducting ($T_c$) transition temperatures and the pseudogap crossover temperature $T^*$ in $YBa_2Cu_3O_{6+y}$. The vertical lines at $p = 0.1$ and $p = 0.25$ mark the positions of two cuprate materials discussed in the text: ortho-II ordered $YBa_2Cu_3O_{6.5}$, located well into the underdoped region, and Tl-2201, well into the overdoped region, respectively.

**b,c**, Distribution of ARPES spectral intensity in one quadrant of the Brillouin zone, measured at $p = 0.1$ (b), on Na-CCOC with $x = 0.10$ (ref.6), and (c), on Tl-2201 at $p = 0.25$ (ref.4). These respectively reveal a truncated Fermi surface made of "Fermi arcs" at $p = 0.10$, and a large, roughly cylindrical and continuous

Fermi surface at $p = 0.25$. The red ellipse in (b) encloses an area $A_k$ that corresponds to the frequency $F$ of quantum oscillations measured in YBCO.

**Figure 2 | Hall resistance of YBa$_2$Cu$_3$O$_{6.5}$ .**

$R_{xy}$ as a function of magnetic field $B$, for sample A, at different temperatures between 1.5 and 4.2 K. The field is applied normal to the CuO$_2$ planes ($B \parallel c$) and the current along the $a$-axis of the orthorhombic crystal structure ($J \parallel a$). **Inset**: Zoom on the data at $T = 2$ K, showing a fitted monotonic background (dashed line).

**Figure 3 | Quantum oscillations in YBCO.**

**a**, Oscillatory part of the Hall resistance, obtained by subtracting the monotonic background (shown in the inset of Fig. 2 for $T = 2$ K), as a function of inverse magnetic field, $1/B$. The background at each temperature is given in Fig. S2.

**b**, Power spectrum (Fourier transform) of the oscillatory part for the $T = 2$ K isotherm, revealing a single frequency at $F = 530 \pm 10$ T, which corresponds to a $k$-space area $A_k = 5.1$ nm$^{-2}$, from the Onsager relation $F = (\, \Phi_0 \,/\, 2\pi^2 \,)\, A_k$ . Note that the uncertainty of 2 % on $F$ is not given by the width of the peak (a consequence of the small number of oscillations), but by the accuracy with which the position of successive maxima in (a) can be determined.

**c**, Temperature dependence of the oscillation amplitude $A$, plotted as $\ln(A\,/\,T)$ vs $T$. The fit is to the standard Lifshitz-Kosevich formula, whereby $A\,/\,T = [\sinh(a\, m^*\, T\,/\, B)]^{-1}$, which yields a cyclotron mass $m^* = 1.9 \pm 0.1$ $m_0$, where $m_0$ is the free electron mass.

**Figure 4 | Fermi surface of YBCO from band structure calculations.**

**a**, Fermi surface of $YBa_2Cu_3O_7$ in the $k_z = 0$ plane (from ref.13), showing the four bands discussed in the main text.

**b,** Fermi surface of ortho-II ordered $YBa_2Cu_3O_{6.5}$ in the $k_z = 0$ plane (from ref.16). In both **a** and **b** the gray shading indicates one quadrant of the first Brillouin zone.

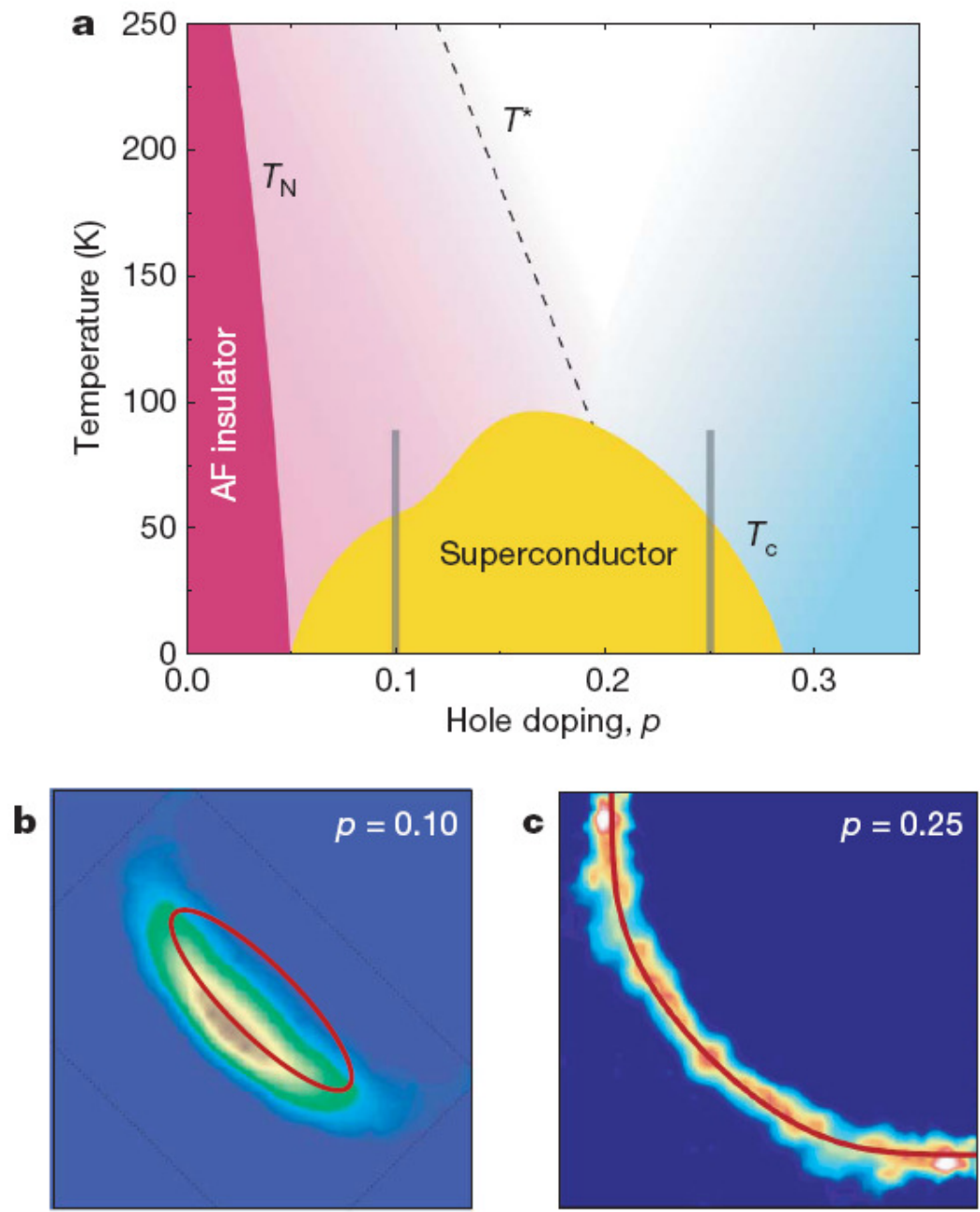

**Figure 1 | Phase diagram of high-temperature superconductors.**

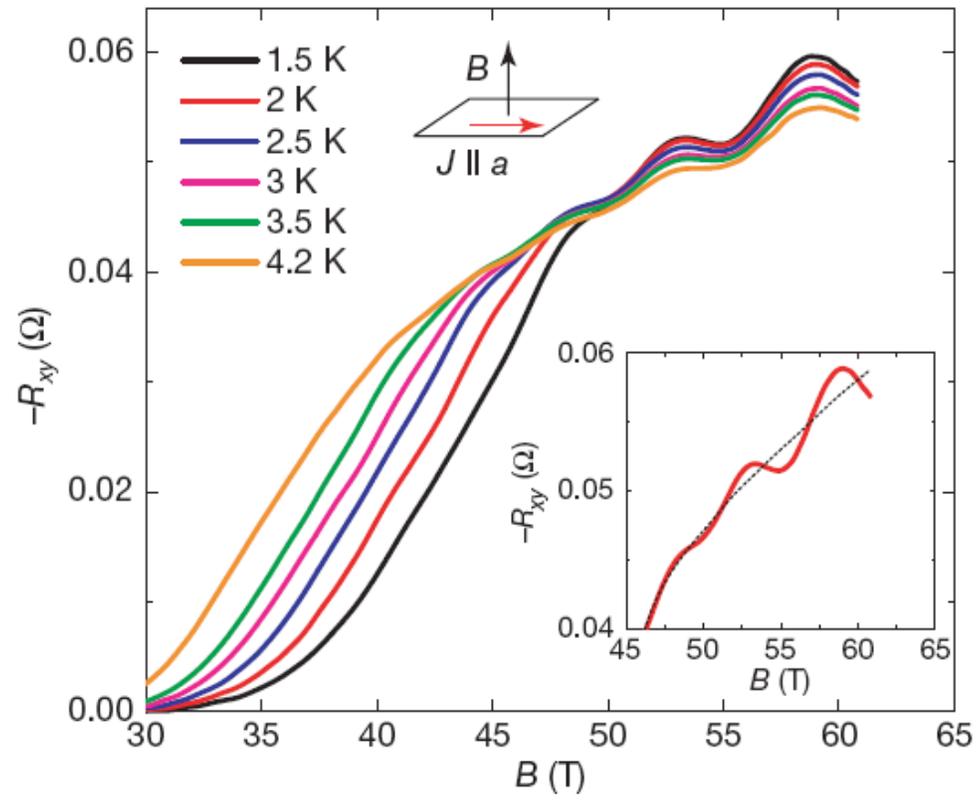

**Figure 2 | Hall resistance of YBa$_2$Cu$_3$O$_{6.5}$ .**

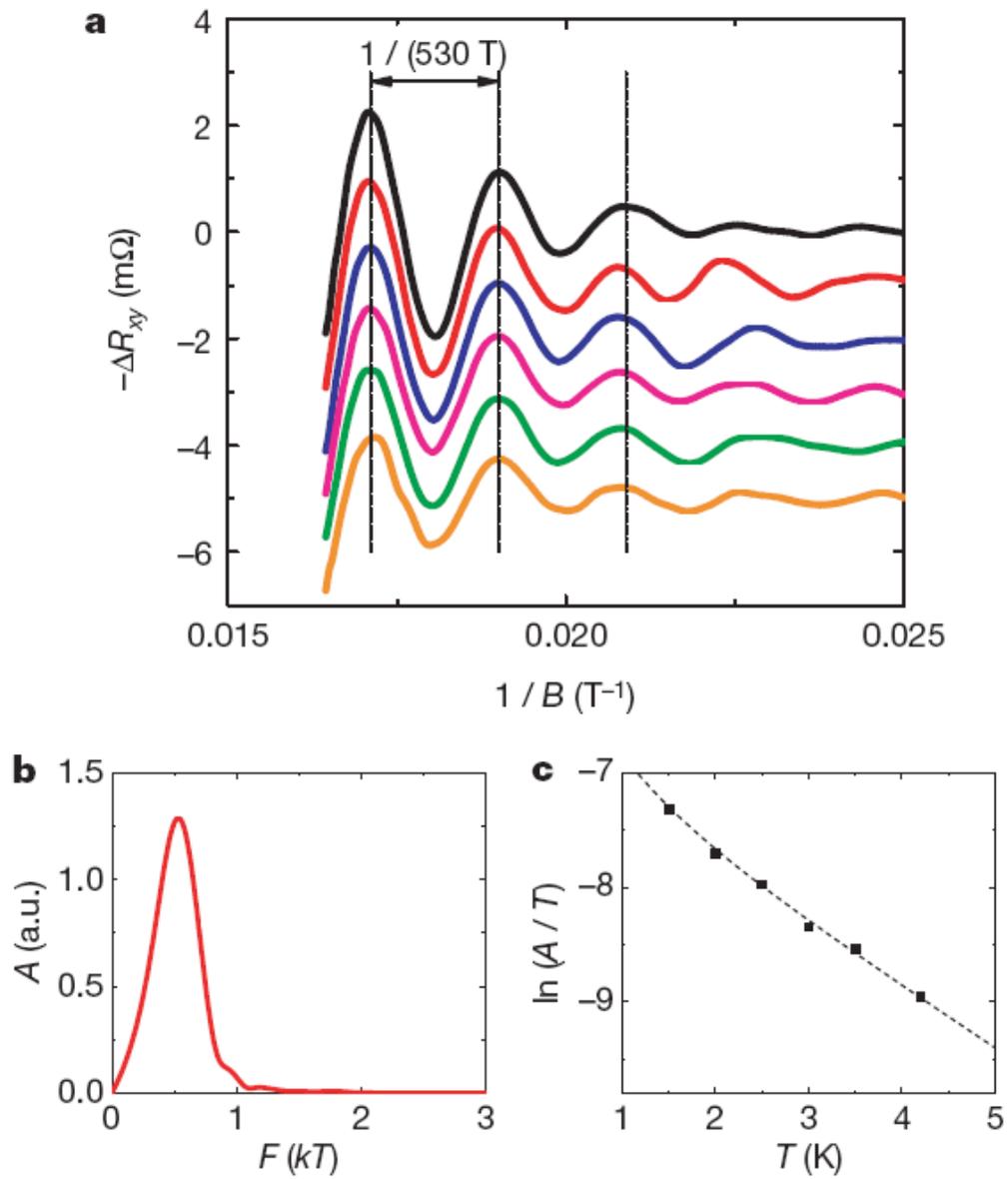

**Figure 3 | Quantum oscillations in YBCO.**

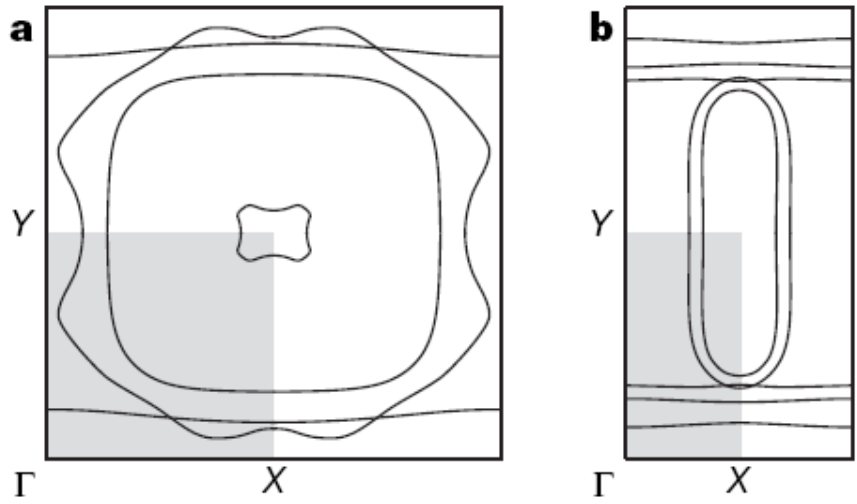

**Figure 4 | Fermi surface of YBCO from band structure calculations.**

# Supplementary information for

# "Quantum oscillations and Fermi surface in an underdoped high temperature superconductor"

Nicolas Doiron-Leyraud [1], Cyril Proust [2], David LeBoeuf [1], Julien Levallois [2], Jean-Baptiste Bonnemaison [1], Ruixing Liang [3,4], D.A. Bonn [3,4], W.N. Hardy [3,4] & Louis Taillefer [1,4]

1 Département de physique & RQMP, Université de Sherbrooke, Sherbrooke, Canada

2 Laboratoire National des Champs Magnétiques Pulsés, UMR CNRS-UPS-INSA 5147, Toulouse, France

3 Department of Physics & Astronomy, University of British Columbia, Vancouver, Canada

4 Canadian Institute for Advanced Research, Toronto, Canada

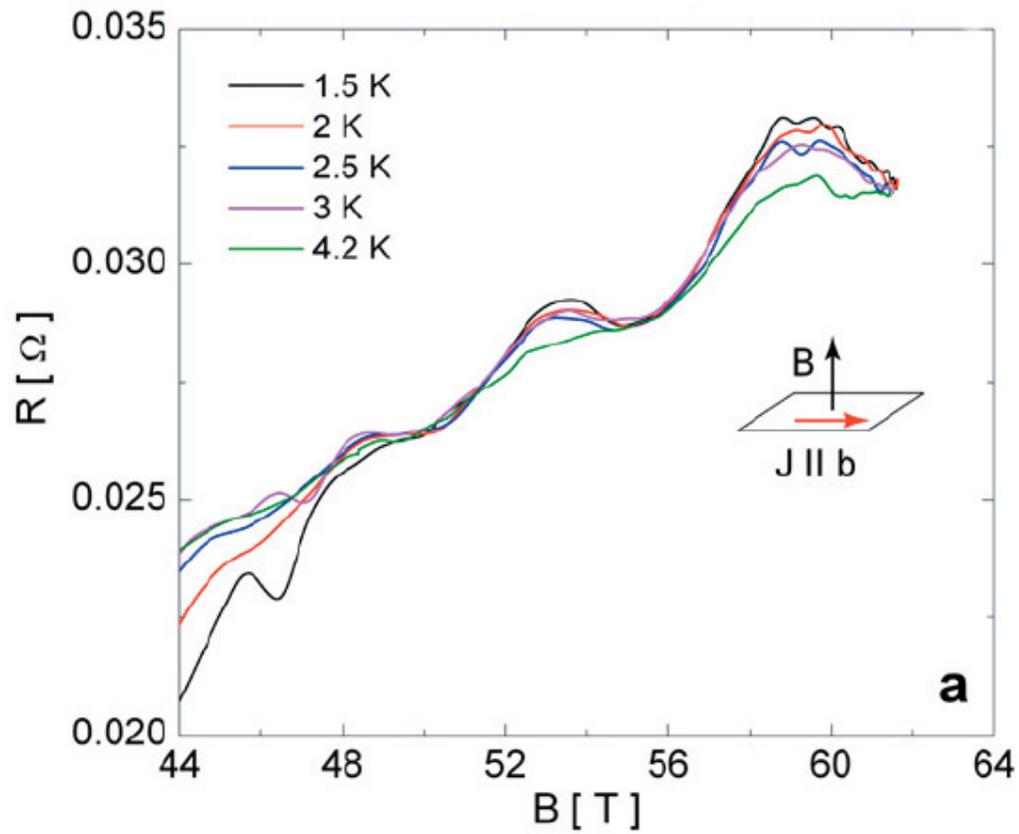

**Fig. S1. a)** Electrical resistance of ortho-II ordered $YBa_2Cu_3O_{6.5}$ (sample B) as a function of magnetic field, at different temperatures $T$ between 1.5 and 4.2 K. The field is applied normal to the $CuO_2$ planes ($B \parallel c$) and the current along the $b$-axis of the ortho-rhombic crystal structure ($J \parallel b$).

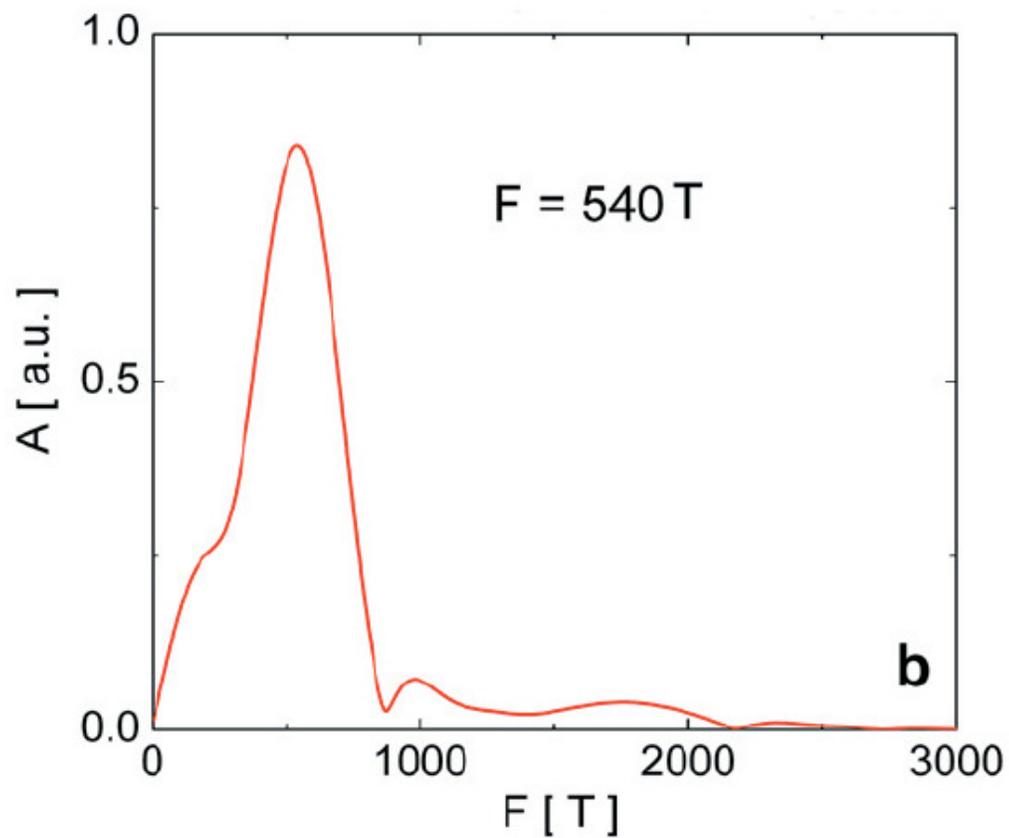

**Fig. S1**. **b)** Power spectrum (Fourier transform) of the oscillatory part for the *T* = 2 K isotherm, revealing a single frequency at *F* = 540 ± 10 T.

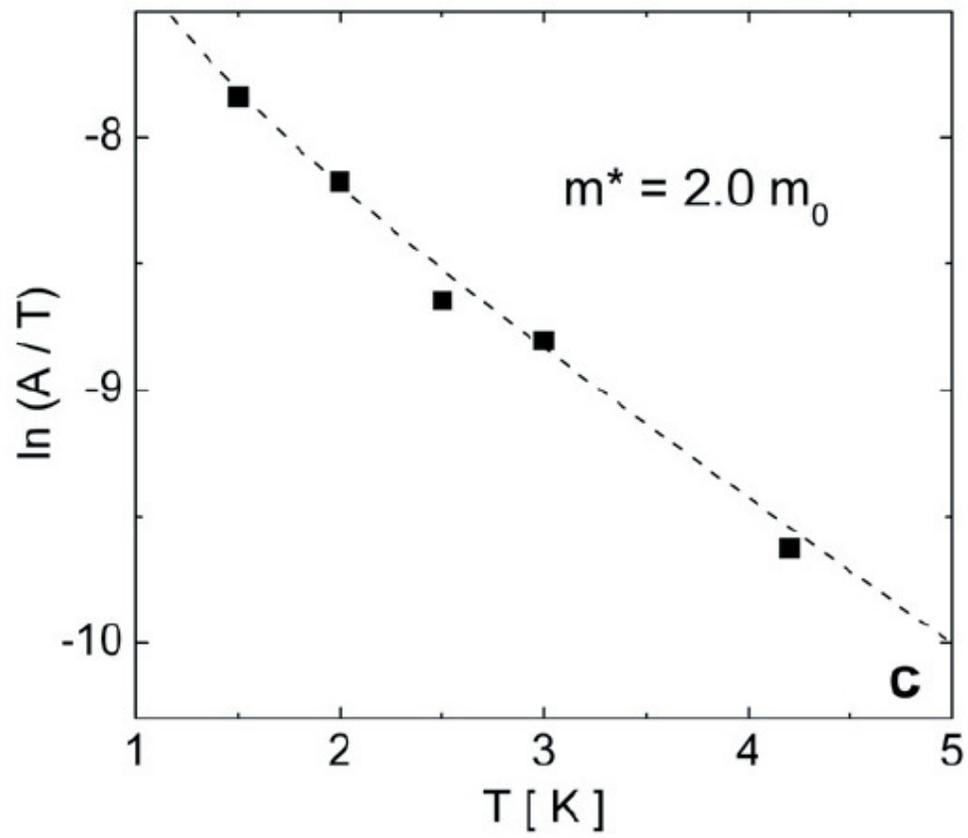

**Fig. S1**.  **c)** Temperature dependence of the oscillation amplitude *A*, plotted as ln(*A* / *T*) vs *T*. The fit to the Lifshitz-Kosevich formula yields a cyclotron mass *m*\* = 2.0 ± 0.1 $m_0$, where $m_0$ is the free electron mass.

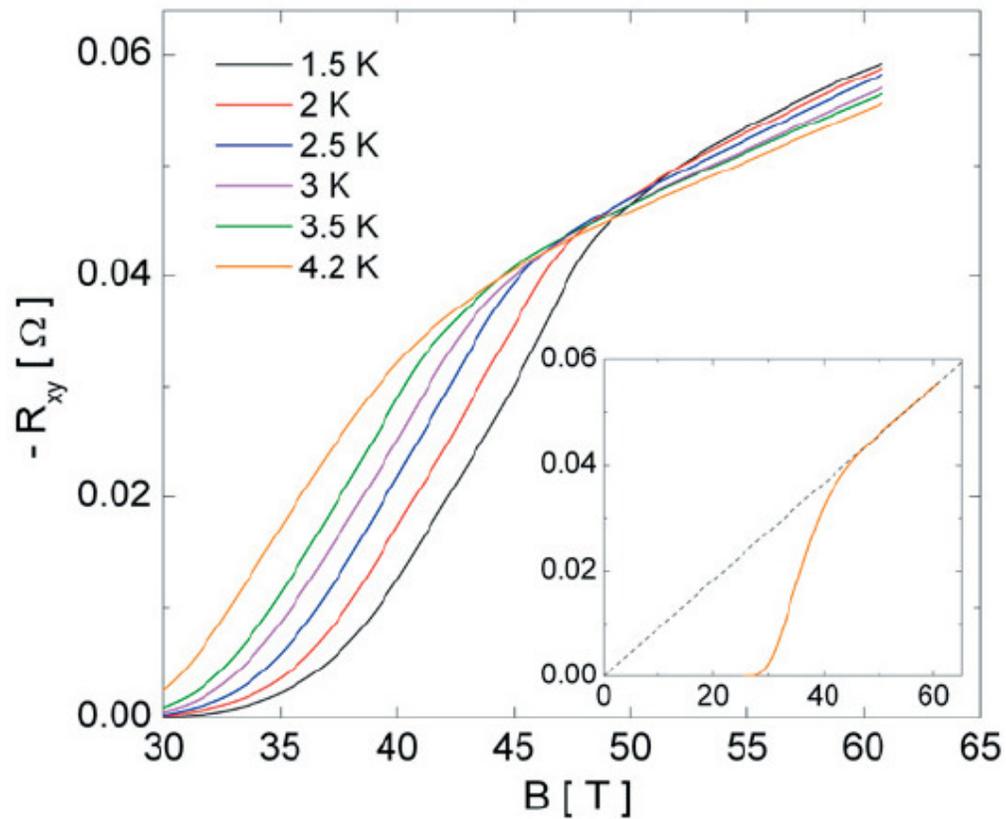

**Fig. S2**. Monotonic part of the Hall resistance of sample A as a function of magnetic field, at different temperatures $T$ between 1.5 and 4.2 K, fitted to the raw data shown in Fig. 2. The oscillatory part displayed in Fig. 3a was obtained by subtracting the monotonic part from the raw data. **Inset**: Zoom on the data at $T$ = 2 K, showing the extrapolation of the monotonic part going to zero as the magnetic field vanishes (dashed line).

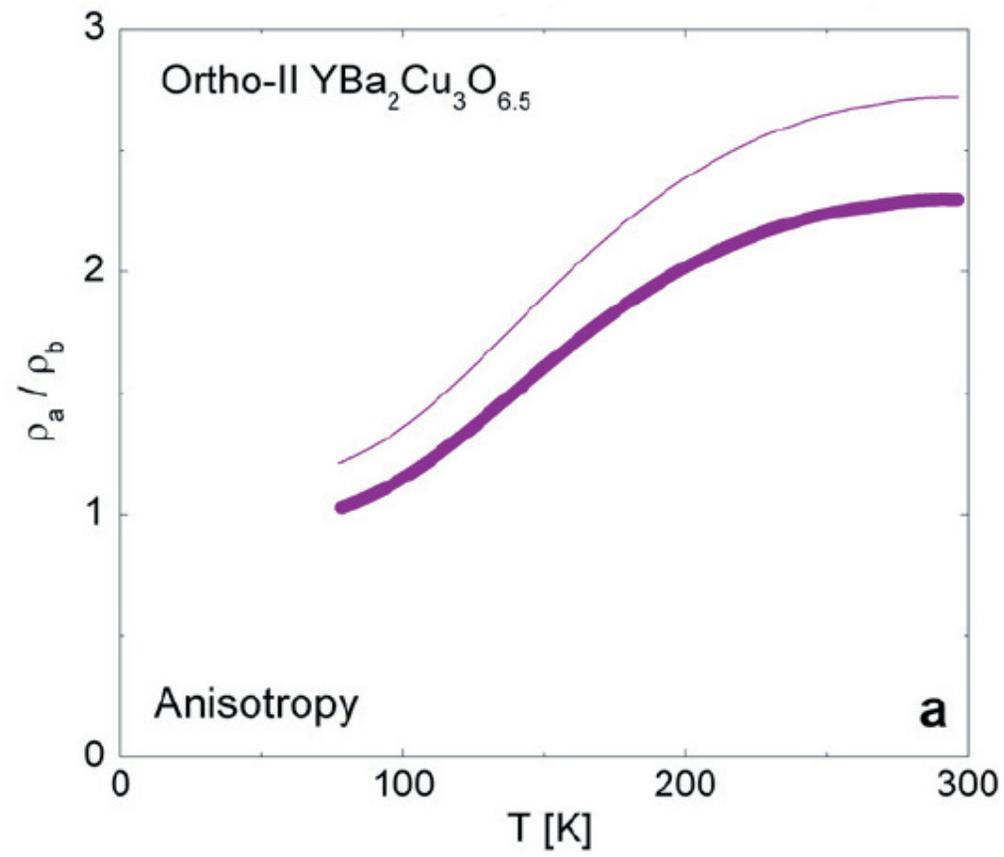

**Fig. S3**. **a)** Ratio of the electrical resistivity measured along the *a* (sample A) and *b* (sample B) axes as a function of temperature.

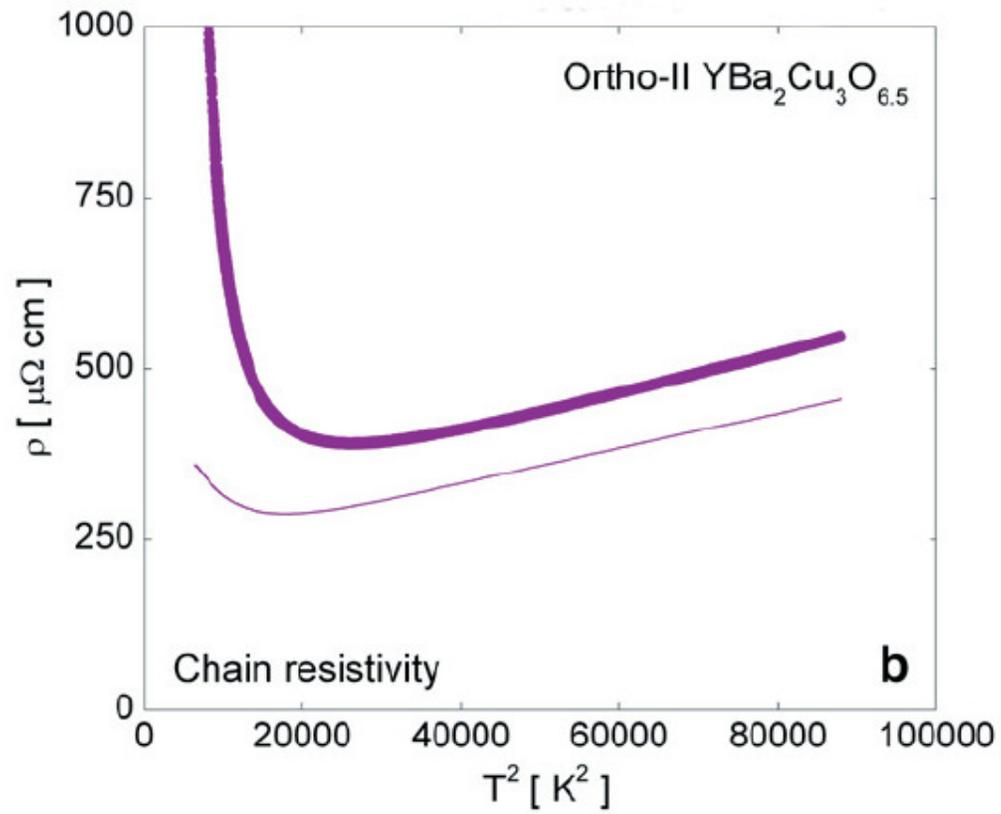

**Fig. S3**. **b)** Resistivity of the chains as a function of temperature. In both panels, the thin line denotes the extremal value allowed by the combined uncertainties on the geometric factor of both samples.

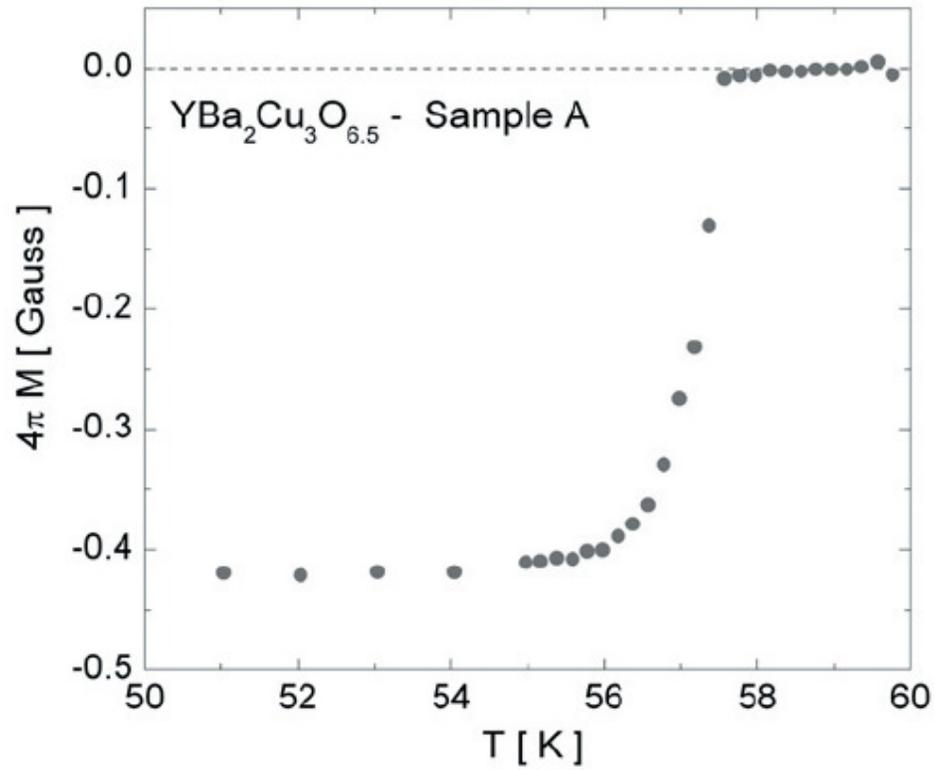

**Fig. S4**. Magnetization of sample A as a function of temperature, showing that the superconducting transition starts sharply at $T_c$ = 57.5 K. This is also precisely where the resistive transition goes to zero.

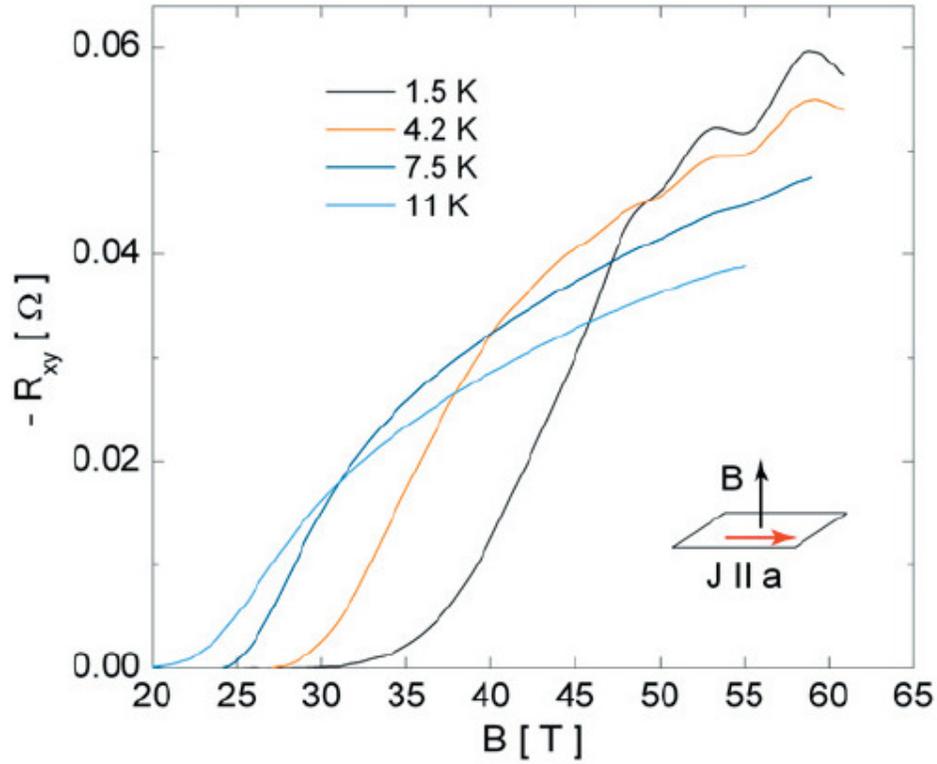

**Fig.**

**S5.** $R_{xy}$ as a function of magnetic field $B$, for sample A, at different temperatures between 1.5 and 11 K. The field is applied normal to the CuO$_2$ planes ($B \parallel c$) and the current along the $a$-axis of the orthorhombic crystal structure ($J \parallel a$). While very weak oscillations are still perceptible at 7.5 K, they are completely absent from the data recorded at 11 K, as expected from thermally damped quantum oscillations.